\documentclass{jaa}

\usepackage{natbib}
\usepackage{graphicx}
\usepackage{float}
\bibpunct{(}{)}{;}{a}{}{,}
\usepackage[colorlinks=true,linkcolor=green,citecolor = blue]{hyperref}%
\makeatletter
\newcommand*{\rom}[1]{\expandafter\@slowromancap\romannumeral #1@}
\makeatother

\def\fm{\hbox{$.\!\!^m$}}
\def\fs{\hbox{$.\!\!^s$}}
\def\sun{\hbox{$\odot$}}                    
\def\degr{\hbox{$^\circ$}}

\newcommand{\apj}{The Astrophy. Journal }
\newcommand{\aj}{AJ}

\newcommand{\mnras}{Monthly Notices of the Royal Astronom. Society}
\newcommand{\pasa}{Publ. Astron. Soc.  Australia}
\newcommand{\aap}{A\&A.}
\newcommand{\aaps}{AAS}

\newcommand{\pasp}{PASP}

\begin{document}

\title{Stellar parameters of the two binary systems: HIP 14075 and HIP 14230}


\author{SUHAIL G. MASDA\textsuperscript{1,2,*}, MASHHOOR A. AL-WARDAT\textsuperscript{3} and J. M. PATHAN\textsuperscript{4}}
\affilOne{\textsuperscript{1}Physics Department, Dr. Babasaheb Ambedkar Marathwada University, Aurangabad-431001, Maharashtra, India.\\}
\affilTwo{\textsuperscript{2}Physics Department, Hadhramout University, PO Box:50511-50512, Mukalla, Yemen.\\}
\affilThree{\textsuperscript{3}Dept. of Physics and Institute of Astronomy and Space Sciences, Al al-Bayt University, PO Box: 130040, Mafraq, 25113 Jordan.\\}
\affilFour{\textsuperscript{4}Physics Department, Maulana Azad College, Aurangabad-431001, Maharashtra, India.}


\twocolumn[{

\maketitle

\corres{suhail.masda@gmail.com}

\msinfo{24 Jan 2018}{YYYY}

\begin{abstract}
We present the stellar parameters of the individual components of the two old close binary systems HIP\,14075 and HIP\,14230 using synthetic photometric analysis. These parameters are accurately calculated based on the best match between the synthetic photometric results within three different photometric systems with the observed photometry of the entire system. From the synthetic photometry, we derive the masses and radii of HIP\,14075 as: $\mathcal{M}^A=0.99\pm0.19\, \mathcal{M_\odot}$, $R_{A}=0.877\pm0.08\, R_\odot$ for the primary and $\mathcal{M}^B=0.96\pm0.15\, \mathcal{M_\odot}$, $R_{B}=0.821\pm0.07\, R_\odot$ for the secondary, and of HIP\,14230 as: $\mathcal{M}^A=1.18\pm0.22\, \mathcal{M_\odot}$, $R_{A}=1.234\pm0.05\, R_\odot$ for the primary and $\mathcal{M}^B=0.84\pm0.12\, \mathcal{M_\odot}$ , $R_{B}=0.820\pm0.05\, R_\odot$ for the secondary, both systems depend on Gaia parallaxes. Based on the positions of the components of the two systems on a theoretical Hertzsprung-
Russell diagram, we find that the age of HIP 14075 is $11.5\pm2.0$\,Gyr and of HIP 14230 is $3.5\pm1.5$\,Gyr. Our analysis reveals that both systems are old close binary systems ($\approx >$ 4 Gyr). Finally, the positions of the components of both systems on the stellar evolutionary tracks and isochrones are discussed.
\end{abstract}

}]


\doinum{}
\artcitid{0000}
\volnum{39}
\year{2018}
\pgrange{1--9}
\setcounter{page}{1}
\lp{9}

\section{Introduction}

The study and analysis of the stellar properties of the close binary systems is a crucial and instrumental in giving a good insight into understanding and improving the knowledge of the formation and evolution of those binaries in many fields of modern astrophysics. Accurate parallaxes of the systems from Gaia mission play a conclusive role in enhancing the values of the abosulte magnitudes and stellar masses of the binary system.

 Determining stellar parameters is stemmed from an accurate analysis of synthetic spectra of the binary systems based on photometric technique. The photometric technique (i.e., colours) of the close binary system is the most extensively used — and most successful technique — in examining the stellar parameters \citep{Al-Wardat4,2018arXiv180203804M}. 
 
 Synthetic photometry is used to estimate the astrophysical parameters
 more accurately through the colour indices \citep{Castelli,Bessell,Straizys,Linnell}. Synthetic photometry is a quantitatively analysis for the synthetic spectral energy distribution (SED) of a binary system which is about modifying stellar parameters
 such that the predicted magnitudes fit the observed ones \citep{Al-Wardat3,Al-Wardat6} and references there in. This procedure leads to precise stellar parameters such as effective temperatures ($T_{\rm eff.}$), radii (R), gravity accelarations (\rm log g), etc.

A combined study of observational measurements and stellar theoretical models is the most powerful manner to analyze binary and multiple systems. So, we follow Al-Wardat's  complex method in analysing such systems \citep{Al-Wardat1}. It is indirect method 
which combines magnitude difference measurements of speckle interferometry,  entire spectral energy distribution (SED) of spectrophotometry and radial velocity measurements (once available), all along with atmospheres modeling (ATLAS9) to determine the individual stellar parameters of the systems \citep{Al-Wardat2012,Al-wardat2014,Al-Wardat5,Masda}.

The selected binaries have distances less than 100 pc and angular separation less than 0.5 arcsec. As a result, they are called visual close binary systems. The followed method to estimate the best stellar parameters for these binaries using synthetic photometry throughout the analysis is primarily a matter of three stages. First, we find the input preliminary parameters of each component to construct the synthetic SED for the entire and individual synthetic SED of the systems using grids of blanketed models (ATLAS9) \citep{Kurucz} which is Al-Wardat's  complex method. Second, we perform synthetic photometry on the synthetic SED to obtain a set of synthetic photometric magnitudes and colour indices for the entire flux and individual components of each system. Third, we compare the results of synthetic photometry with the observed photometry of the binary systems. The latter stage is extremely important to tell us about accurate physical parameters which are tested with the best match between synthetic photometry and the observed photometry of each system.

The HIP 14075 and HIP 14230 are well-known as visual close binary systems in the solar neighborhood located at a distance of  60.17$\pm$0.001 and 37.99 $\pm$ 0.001 pc \cite{2018yCat.1345....0G}, respectively.

\cite{Balega2005} studied the orbits of those binaries and concluded that they have late G- or early K-type main sequence components. They found that the sum masses of the  HIP 14075 system were $2.03\pm0.58\mathcal{M_\odot}$ using old \textit{Hipparcos} parallax of $15.17\pm1.44$ mas and of the HIP 14230 system were $1.47\pm0.26\mathcal{M_\odot}$ using old \textit{Hipparcos} parallax of $29.62\pm1.09$ mas. They stated that the masses of these systems are derived with low accuracy and high \textit{Hipparcos} parallax error. Their results were obtained by using orbital solutions of each system.

The aims of the present work are to find the accurate stellar parameters of the systems that
can achieve the best match between the observed
magnitudes and colour indices of the entire system 
with the entire synthetic ones and to employ the Gaia parallaxes of each system in our study.   
\section{Observational photometric data}
In our study, we depend on observational photometric data which are taken from reliable different sources such as Hipparcos \citep{ESA}, Str\"{o}mgren \citep{Hauck} and TYCHO
catalogues \citep{Hog}. These data are used as reference and comparison with synthetic photometric results to get the best stellar parameters of the systems.
Table~\ref{tabl0} contains fundamental parameters and the observed photometric data for HIP\,14075 and HIP\,14230 from SIMBAD database, NASA/IPAC, Str\"{o}mgren, the Hipparcos and TYCHO
catalogues and Table~\ref{taba1} shows the magnitudes difference $\rm \Delta m$ between the components of both
systems along with filters used in the observation expressed in nm and reference for each value from Fourth Catalog of Interferometric Measurements of Binary Stars (INT4)~\footnote{http://www.usno.navy.mil/USNO/astrometry/optical-IR-prod/wds/int4}

\begin{table}[htb]
	\centering
	\caption{Fundamental parameters and observed photometric data for two systems HIP 114075 and HIP 14230.} \label{tabl0}
	\begin{tabular}{cccc}\hline
		Property & HIP 14075 & HIP 14230  & Ref.  \\
		\hline			\noalign{\smallskip}
		$\alpha_{2000}$ $^{a}$  & $03^h 01^m 22\fs687$ & $03^h 03^m 28\fs656$ & simbad\\
		$\delta_{2000}$ $^{b}$  & $+06\degr14' 56.''85$	& $+23\degr03' 41.''33$ & -\\
		Sp. Typ.  &  G5  & G0 & -
		\\
		E(B-V)& $ 0.099\pm0.01 $ & $ 0.17\pm0.004 $& $c$ 
		\\		
		$A_v$   &  $0\fm30$  &$0\fm52$ &$c$
		\\
		Gaia DR2  (mas) &    $16.62\pm0.29$   &$26.32\pm0.70$ & $d$\\
			$V_J$  & $8\fm82$  & $7\fm09$& $e$
			\\
			$(V-I)_J$& $0\fm79\pm0.01$ & $0\fm69\pm0.01$ & -
			\\				
			$(B-V)_J$& $0\fm741\pm0.02$ & $0\fm624\pm0.006$ & -
			\\
			$(b-y)_S$& $0\fm44\pm0.002$ & $0\fm39\pm0.003$ & $f$
			\\
			$(v-b)_S$& $0\fm69\pm0.004$ & $0\fm60\pm0.008$ & -\\
			$(u-v)_S$& $0\fm94\pm0.013$ & $0\fm90\pm0.014$ & -\\
									
			$B_T$  &   $9\fm73\pm0.02$  & $7\fm81\pm0.007$ & $ g$
			\\			
			$V_T$ &   $8\fm91\pm0.02$   & $7\fm12\pm0.007$ & -
				\\
		\hline
	\end{tabular}
	\\
	\tablenotes{\textbf{Notes.} $^{a}$ Right Ascention and $^{b}$ Declination.\\
 $^{c}$ The interstellar reddening is: $A_v$=3.1 $E(B-V)$ where $E(B-V)$ is towards the direction of these stars \citep{Schlafly}, $^d$ \citep{2018yCat.1345....0G}, $^e$  \citep{ESA}, $^f$ \citep{Hauck} and $^g$ \citep{Hog}}
\end{table}

\begin{table}[h]
	\begin{center}
		\caption{Speckle interferometric magnitude differences and
			Hipparcos $ \Delta H_{Hip}$ measurements of both systems, along with filters used to obtain the observations.}
		\label{taba1}
		\begin{tabular}{c|cccc}
			\noalign{\smallskip}
			\hline
		HIP	&$\triangle m $& {$\sigma_{\Delta m}$}& Filter ($\lambda/\Delta\lambda$)& Ref.  \\
			\hline
	14075	&	$0\fm44$ &   0.29  & $V_{Hip}:545nm/30$&  1   \\
			
		&	$0\fm00$ &   0.29  & $800nm/60$&  2   \\		
		&	$0\fm10$ &   0.13  & $545nm/30$&  3    \\						
		&	$0\fm17$ &  0.16  & $545nm/30$&  4   \\
			
		&	$0\fm00$ &   0.22  &$600nm/30 $&  5     \\
		&	$0\fm00$ &  0.22  & $600nm/30$ &   5   \\
		&	$0\fm00$ &  0.22 &$600nm/30$  &  6   \\
			
		&	$0\fm25$ &   -  &$550nm/40 $&  7    \\
		&	$0\fm00$ &  0.12  & $550nm/40 $ &  8  \\
		&	$0\fm40$ &  - &$551nm/22$  &  9  \\
			
			\hline
	14230	&	$1\fm72$ &   0.25  &$V_{Hip}:545nm/30$&  1   \\
		&	$1\fm72$ &   0.04  & $545nm/30$& 3    \\		
		&	$1\fm74$ &   0.05  & $545nm/30$&  2  \\		
		&	$1\fm72$ &   0.03  & $545nm/30$& 3    \\
		&	$1\fm74$ &  0.03  & $545nm/30$& 4   \\
		&	$1\fm31$ &   0.08  &$648nm/41 $& 10    \\
		&	$1\fm53$ &   0.04  &$600nm/30 $& 5    \\
		&	$1\fm55$ &  0.02  & $600nm/30 $ & 5   \\
		&	$1\fm55$ &  0.01 &$600nm/30 $  & 6  \\
		&	$1\fm73$ &  0.06  & $545nm/30 $ & 11     \\	\hline
		\end{tabular}
		\\
		\tablenotes{
			$^1$\citep{ESA},
			$^2$\citep{Balega2002},
			$^3$\citep{Pluzhnik2005},
			$^4$\citep{Balega2004},
			$^5$\citep{Balega2006b},
			$^6$\citep{Balega2005},
			$^7$\citep{Horch2010},
			$^8$\citep{Horch2012},
			$^9$\citep{Tokovinin2010},
			$^{10}$\citep{Horch2004},
			$^{11}$\citep{Balega2007}}.
	\end{center}
\end{table}

\section{Analysis methods}
\subsection{HIP 14075 (HDS 385)}\label{10}
In order to follow Al-Wardat's  complex method  in deriving the stellar
parameters of the individual components of the binary system HIP 14075, it is bound to know the magnitude differences measured by speckle interferometry between the components of the system and the Gaia parallax for the system \citep{2018yCat.1345....0G}. Therefore, we determined the visual magnitude difference of the system as $\triangle m=0\fm 24\pm0.12$ between the two components as the average for three $\triangle m$ measurements given in Table~\ref{taba1} under the speckle V-band filter $545nm/30$.

Using a magnitude difference between the components 
of the system $\Delta m$, visual magnitude $m_{v}$ ($V_{J}$) and a revised distance of Gaia, we readily compute the apparent and absolute visual
magnitudes for components of the system using the following equations \citep{Heintz} (see p.28):
{\begin{eqnarray}\centering
	\ m_v^A=m_v+2.5\log(1+10^{-0.4\triangle m}),
	\label{eq11}
	\end{eqnarray}
	\begin{eqnarray}
	\centering
	\label{eq22}
	\ m_v^B=m_v^A+{\triangle m}.
	\end{eqnarray}} And
\begin{eqnarray}
\label{eq3}
\ M_V=m_v+5-5\log(d)-A_v
\end{eqnarray}

\noindent which give: $m_v^A=9\fm46\pm0.05 , M_V^A=5\fm56\pm0.06$  and  $ m_v^B=9\fm70\pm0.13, M_V^B=5\fm80\pm0.14$ for the primary and secondary components of the system, respectively. In both stars, we assumed no absorption ($\ A_{v}=0$) because the studied binary systems here are nearby stars \citep{Balega2006b}.

In order to calculate the errors of the apparent and absolute visual magnitudes of the components of the system in Eqs.~\ref{eq11},~\ref{eq22}~\&\ref{eq3}, we use the following sample relations:

\begin{eqnarray} 
\label{eq33}
\sigma_{m^A_{v}} =\pm \sqrt{\sigma_{m_{v}}^2+(\frac{10^{-0.4\triangle m}}{1+10^{-0.4\triangle m}})^2\sigma_{\triangle m}^2},
\end{eqnarray}

\begin{eqnarray}
\label{eq331}
\sigma_{m^B_{v}} =\pm \sqrt{\sigma_{m^A_{v}}^2+\sigma_{\triangle m}^2}.
\end{eqnarray}

And

\begin{eqnarray} 
\label{eq341}
\sigma_{M^{*}_{V}} =\pm \sqrt{\sigma_{m^{*}_{v}}^2+(\frac{5 \log e}{\pi_{Hip}})^2\sigma_{\pi_{Hip}}^2 +\sigma_{A_{v}}^2}.
\end{eqnarray}
Here, the error of visual magnitude $ \sigma_{m_{v}} $ is very tiny and not given in data of SIMBAD (Table~\ref{tabl0}), because of this it can be neglected in Eq.~\ref{eq33} and  $^*$ indicates A and B components of the system in Eq.\ref{eq341}.

Based on the above estimated absolute magnitudes ($M_{V}$) of the individual components of the system  and their relations with effective temperatures ($T_{\rm eff.}$) in addition to Tables \citep{Lang,Gray} and the following equations:
\begin{eqnarray}
\label{eq8}
\log(R/R_\odot)= 0.5 \log(L/L_\odot)-2\log(T_{\rm eff.}/T_\odot)\\
\label{eq5}
\log g = \log(M/M_\odot)- 2\log(R/R_\odot) + 4.43,
\end{eqnarray}
 the input preliminary parameters of the system estimate as follows: $T_{\rm eff.}=5500K$, log g = 4.48, $R=0.85R_\odot$ for the primary component and $T_{\rm eff.}=5380K$, log g = 4.55, $R=0.77R_\odot$ for the secondary component. Here
 $T_\odot$ was taken as $5777\rm{K}$.

In order to further improve the above primary parameters, we need to construct the synthetic spectral energy distribution (SED) of the system based on input parameters  and on grids of blanketed models (ATLAS9) \citep{Kurucz}.

Hence, the entire synthetic SED at Earth of the binary system, which is connected to the energy flux of the individual components, is computed using the following equation:

\begin{eqnarray}
	\label{eq7}
	F_\lambda  = (R_{A} /d)^2(H_\lambda ^A + H_\lambda ^B \cdot(R_{B}/R_{A})^2) ,
\end{eqnarray}

\noindent
where $ R_{A}$ and $ R_{B}$ are the radii of the primary and secondary components of the system in solar units, $H_\lambda ^A $ and  $H_\lambda ^B$ are the fluxes at the surface of the star and $F_\lambda$ is the flux for the entire SED of the system above  the Earth's atmosphere which is located at a distance d (pc) from the system. 

If the effective temperatures and gravity accelarations of each component are available of the binary system in addition to radii and distance of the binary system, we get the  $H_\lambda ^A $ and  $H_\lambda ^B$ data as fluxes and wavelengths using blanketed models (Atlas9) \citep{Kurucz}. The combination between $H_\lambda ^A $ and  $H_\lambda ^B$ yields the entire flux of the binary system.

Owing to the lack of knowledge of the observed spectrum of the system, we use a new technique which completely performs the same role in analysing the binary system. This technique is primarily dependent on the results of Al-Wardat's method for the same binary. Therefore, in order to examine those stellar parameters which were used to construct preliminary synthetic SED of the system, we should use synthetic photometry of the system.

\subsubsection{Synthetic photometry} \label{12}
$\\$$\\$
The stellar parameters are mainly dependent on the best match between the observed coulors indices and magnitudes of the entire system with the entire synthetic SED of the system. Therefore, the entire and individual synthetic magnitudes and coulors indices of the binary system are calculated by integrating the model fluxes over each bandpass of the system calibrated to the reference star (Vega) using the following equation \citep{Maiz2007,Al-Wardat2012}:

\begin{equation}\label{15}
m_p[F_{\lambda,s}(\lambda)] = -2.5 \log \frac{\int P_{p}(\lambda)F_{\lambda,s}(\lambda)\lambda{\rm d}\lambda}{\int P_{p}(\lambda)F_{\lambda,r}(\lambda)\lambda{\rm d}\lambda}+ {\rm ZP}_p\
\end{equation}
where $m_p$ is the synthetic magnitude of the passband $p$, $P_p(\lambda)$ is the dimensionless sensitivity function of the passband $p$, $F_{\lambda,s}(\lambda)$ is the synthetic SED of the object and $F_{\lambda,r}(\lambda)$ is the SED of Vega.  Zero points (ZP$_p$) from ~\citep{Maiz2007} (and references there in) were adopted.

Many attempts were carried out in revising the stellar parameters to obtain the best-fit between the observed magnitudes and  colour indices of the entire system  with the
entire synthetic ones.

Hence, the final findings of the calculated magnitudes and color  indices (Johnson: $U$, $B$, $ V$, $R$, $U-B$, $B-V$, $V-R$; Str\"{o}mgren: $u$, $v$, $b$,
$y$, $u-v$, $v-b$, $b-y$ and Tycho: $B_{T}$, $ V_{T}$, $B_{T}-V_{T}$) of the entire
synthetic system and individual components of the system HIP 14075, in different photometrical systems,  are shown in Table~\ref{s1}.

By following all steps and iterative method, the accurate stellar parameters of the components of the system are those which led to the best match between the observed magnitudes and  colour indices of the entire system (see Table~\ref{tabl0}) with the
entire synthetic ones (see Table~\ref{s1}). These parameters ($T_{\rm eff.}$, $\log\rm g$, $R$ and $d$ ) are listed in Table~\ref{tablef1} and displayed in Fig.~\ref{fig1}.

Based on the ultimate radii and effective temperatures of the system, we calculated the stellar luminosities and bolometric magnitudes along with their errors which are listed in Table~\ref{tablef1}. 

Depending on the final calculated parameters of the individual comonents and Tables of ~\cite{Lang} and \cite{Gray}, the spectral type of HIP 14075 A is
found to be G4.5V and HIP 14075 B to be G6.5V.

	\subsection{HIP 14230 (HDS 389)}
	Following the same procedures applied to calculate preliminary parameters in the preceding section \ref{10}, taking into account the visual magnitude $\rm m_v=7^{\rm m}.09$ given in Table~\ref{tabl0} in addition to magnitude difference  $\Delta m=1^{\rm m}.73 \pm0.05$ as average values for six $\triangle m$ measurements given in Table~\ref{taba1} under  speckle filters for V-band filters 545nm/30 nm, we attain the apparent magnitudes as: $\rm m_v^A=7^{\rm m}.29\pm0.01$ and $\rm m_v^B=9^{\rm m}.02\pm0.05$ for the primary and secondary components, respectively. Based on the above estimated apparent magnitudes and on Gaia parallex, all these values lead to the absolute magnitudes as: $\rm M_V^A=4^{\rm m}.39\pm0.06$ and $\rm M_V^B=6^{\rm m}.12\pm0.08$ for the primary and secondary components, respectively.

	Now then, as formerly mentioned regarding the input parameters of the system and on the strength of that, we obtain:
	$T_{\rm eff.}=5950K$, log g = 4.35, $R=1.12R_\odot$ for the primary component and $T_{\rm eff.}=5225K$, log g = 4.54, $R=0.78R_\odot$ for the secondary component.
	
	In order to enhance the above obtained input parameters, we should construct the entire synthetic SED  and its individual components of the binary system.

	For the sake of examining the stellar parameters which were used to construct the synthetic  SED, the synthetic photometry should be used  for the system.
	
	\subsubsection{Synthetic photometry}
	$\\$$\\$
	Following the same procedures applied in the preceding section \ref{12}, the final entire and individual synthetic magnitudes within three different photometric systems: Johnson-Cousins ($U$, $B$, $ V$, $R$, $U-B$, $B-V$, $V-R$), Str\"{o}mgren ($u$, $v$, $b$,
	$y$, $u-v$, $v-b$, $b-y$) and Tycho ($B_{T}$, $ V_{T}$, $B_{T}-V_{T}$) of the close binary system HIP 14230 are accurately calculated using the equation \ref{15} and listed in Table~\ref{s1}.

	Hence, the best match was achieved at set of the stellar parameters which are listed in Table~\ref{tablef1} and displayed in Fig.~\ref{fig2}.

	These stellar parameters represent the real parameters for the system depending on real photometric analysis. Thus,  
	the final individual stellar luminosities and bolometric magnitudes of the binary system are calculated and listed in Table~\ref{tablef1}.

	Finally, we put the components of two close binary systems, HIP 14075 and HIP 14230, in the theoretical Hertzsprung-Russell (H-R)
	diagram, log $\rm L/\rm L_\odot$ versus log $T_{\rm eff}$, from \cite{Girardi2000b} (see Fig.\ref{a25}), completed with the
	isochrones computed by \cite{Girardi2000a}  (see Fig.\ref{a26}). The positions of the components of the stars in these diagrams lead to theoretical estimates of their masses and ages, and give explicit ideas of their evolutionary status.
	
	\section {Results and discussion}\label{4}
	
	Table~\ref{s1} shows  synthetic photometry magnitudes and colour indices of the entire and individual synthetic SED of the HIP\,14075 and HIP\,14230 binary systems within three different photometric systems: Johnson-Cousins, Str\"{o}mgren and Tycho.
	
	\begin{table*}
		\small
		\begin{center}
			\caption{ Magnitudes and color indices  of the entire synthetic system and individual components of the two systems HIP 14075 and HIP 14230.}
			\label{s1}
			\begin{tabular}{ccccc|ccc}
				\noalign{\smallskip}
				\hline
				\noalign{\smallskip}
				&  & \multicolumn{3}{c|}{HIP\,14075} & \multicolumn{3}{c}{HIP\,14230} \\
				\cline{3-5} \cline{6-8}
				&  & Entire synth. & HIP 14075 A& HIP 14075 B & Entire synth.  & HIP 14230 A & HIP 14230 B \\
				System	&   Filter  & (mag)&   (mag)    &     (mag)  &     (mag) &     (mag) &     (mag)      \\
				\hline
				\noalign{\smallskip}
				Johnson-Cousins & $U$  & $9.85\pm0.03$  & 10.45  & 10.77 & $7.85\pm0.03$   & 7.96  & 10.35\\
				& $B$     & $9.56\pm0.03$  &  10.19 & 10.46 & $7.71\pm0.03$   &  7.87 & 9.87 \\
				& $V$                   & $8.82\pm0.03$  &  9.46 &  9.70 & $7.09\pm0.03$   &  7.29 &  9.02\\
				& $R$                       & $8.43\pm0.03$   &  9.07 & 9.29 & $6.75\pm0.03$   &  6.97 & 8.56 \\
				&$U-B$                  & $0.28\pm0.04$   & 0.26  & 0.31 & $0.14\pm0.04$   & 0.09  & 0.48 \\
				&$B-V$                  & $0.74 \pm0.04$  &  0.73 &  0.76 & $0.62\pm0.04$   &  0.58 &  0.85 \\
				&$V-R$                  & $0.40 \pm0.04$  &  0.39 & 0.40 & $0.34\pm0.04$   &  0.32 & 0.46 \\\hline
				\noalign{\smallskip}
				Str\"{o}mgren        & $u$ & $10.99\pm0.03$   & 11.60  &  11.91  & $9.01\pm0.03$   & 9.12  &  11.50 \\
				& $v$                   & $9.96\pm0.03$   & 10.58  & 10.86 & $8.05\pm0.03$   & 8.19  & 10.34 \\
				& $b$                   & $9.23\pm0.03$   & 9.86  &  10.11 & $7.44\pm0.03$   & 7.62  &  9.47 \\
				&  $y$                  & $8.79\pm0.03$   & 9.43  & 9.66   & $7.06\pm0.03$   & 7.26  & 8.98 \\
				&$u-v$                  & $1.03\pm0.04$   & 1.02  & 1.05 & $0.96\pm0.04$   & 0.94  & 1.16 \\
				&$v-b$                  & $0.73\pm0.04$   & 0.71  & 0.75 & $0.61\pm0.04$   & 0.57  & 0.87\\
				&$b-y$                  & $0.44\pm0.04$   & 0.43  & 0.45 & $0.38\pm0.04$   & 0.36  & 0.49 \\\hline
				\noalign{\smallskip}
				Tycho       &$B_T$      & $9.75\pm0.03$   & 10.38 & 10.65   & $7.86\pm0.03$   & 8.01 & 10.10 \\
				&$V_T$                  & $8.90\pm0.03$   & 9.54 & 9.78 & $7.16\pm0.03$   & 7.36 & 9.11  \\
				&$B_T-V_T$              & $0.85\pm0.04$   & 0.84 & 0.87 & $0.70\pm0.04$   & 0.65 & 0.99\\
				\hline
			\end{tabular}
		\end{center}
	\end{table*}

	The comparison between  the entire synthetic magnitudes and color indices (Table \ref{s1}) with the observed ones (Table \ref{tabl0}) shows a most good agreement within three photometric systems: Johnson-Cousins, Str\"{o}mgren and Tycho. This comparison was crucial and led to the most important indication for the reliability of the calculated stellar parameters of two old close binary systems, HIP 14075 and HIP 14230, listed in Table~\ref{tablef1}. In addition to that, the magnitude differences $\triangle m$ of both systems from the synthetic photometry ($\triangle m= V^{B}_{J}-V^{A}_{J}$) (Table \ref{s1}) are found to be exactly similar those from the observed ones (Table \ref{taba1}).
	
		\begin{table*}
			\small
			\begin{center}
				\caption{The ultimate  stellar parameters of the components of the systems HIP 14075 and HIP 14230.} \label{tablef1}
				\begin{tabular}{ccc|cc}
					\noalign{\smallskip}
					\hline
					\noalign{\smallskip}
					&\multicolumn{2}{c|}{HIP 14075}&\multicolumn{2}{c}{HIP 14230}\\
					\cline{2-3}\cline{4-5}
					\noalign{\smallskip}
					Parameter	& HIP 14075 A & HIP 14075 B & HIP 14230 A & HIP 14230 B \\
					\hline
					\noalign{\smallskip}
					$\rm T_{\rm eff.}$ {$\pm$ $\sigma_{\rm T_{\rm eff}}$}[K] & $5670\pm100$ & $5580\pm100$ & $6140\pm100$ & $5300\pm100$\\
					R {$\pm$ $\sigma_{\rm R}$}  [R$_{\odot}$] & $0.877\pm0.08$ & $0.821\pm0.07$ & $1.234\pm0.05$ & $0.820\pm0.05$ \\
					$\log\rm g$ {$\pm$ $\sigma_{\rm log g}$} [cgs]&$4.45\pm0.12$ & $4.50\pm0.11$ &$4.30\pm0.08$ & $4.55\pm0.09$\\
					$\rm L $ {$\pm$ $\sigma_{\rm L}$} [$\rm L_\odot$] & $0.71\pm0.06 $  & $0.59\pm0.09$ & $1.94\pm0.08 $  & $0.48\pm0.10$\\
					$\rm M_{bol}$ {$\pm$ $\sigma_{\rm M_{bol}}$}  [mag.] &  $5\fm12\pm0.06$ & $5\fm32\pm0.14$ &  $4\fm03\pm0.06$ & $5\fm55\pm0.08$\\
					$\mathcal{M}$ $^{1}$ {$\pm$ $\sigma_{\rm \mathcal{M}}$}  [$\rm \mathcal{M_\odot}$]&  $0.99 \pm0.19$ & $0.96 \pm0.15$ &  $1.18 \pm0.22$ & $0.84 \pm0.12$\\
					Sp. Type$^{2}$  &  G4.5V & G6.5V &  F8V & K0V\\
					\hline\noalign{\smallskip}
					\multicolumn{1}{c}{Age $^{3}$ [Gyr]}& \multicolumn{2}{c}{ $11.5\pm 2.0$} & \multicolumn{2}{c}{ $3.5\pm 1.5$}\\
					\hline\noalign{\smallskip}
				\end{tabular}\\
				\tablenotes{\textbf{ Notes.} $^{1}${Depending on the evolutionary tracks of~\citep{Girardi2000b} (Fig.~\ref{a25})},\\
					$^{2}${Using the tables of~\citep{Lang,Gray}.\\
						$^{3}${Depending on the the isochrones of
							different metallicities of~\citep{Girardi2000a} ( Figs.~\ref{a27}\&~\ref{a28}}).}}
			\end{center}
		\end{table*}
	
\begin{figure*}
    	\centering
    	\includegraphics[angle=0,width=12cm]{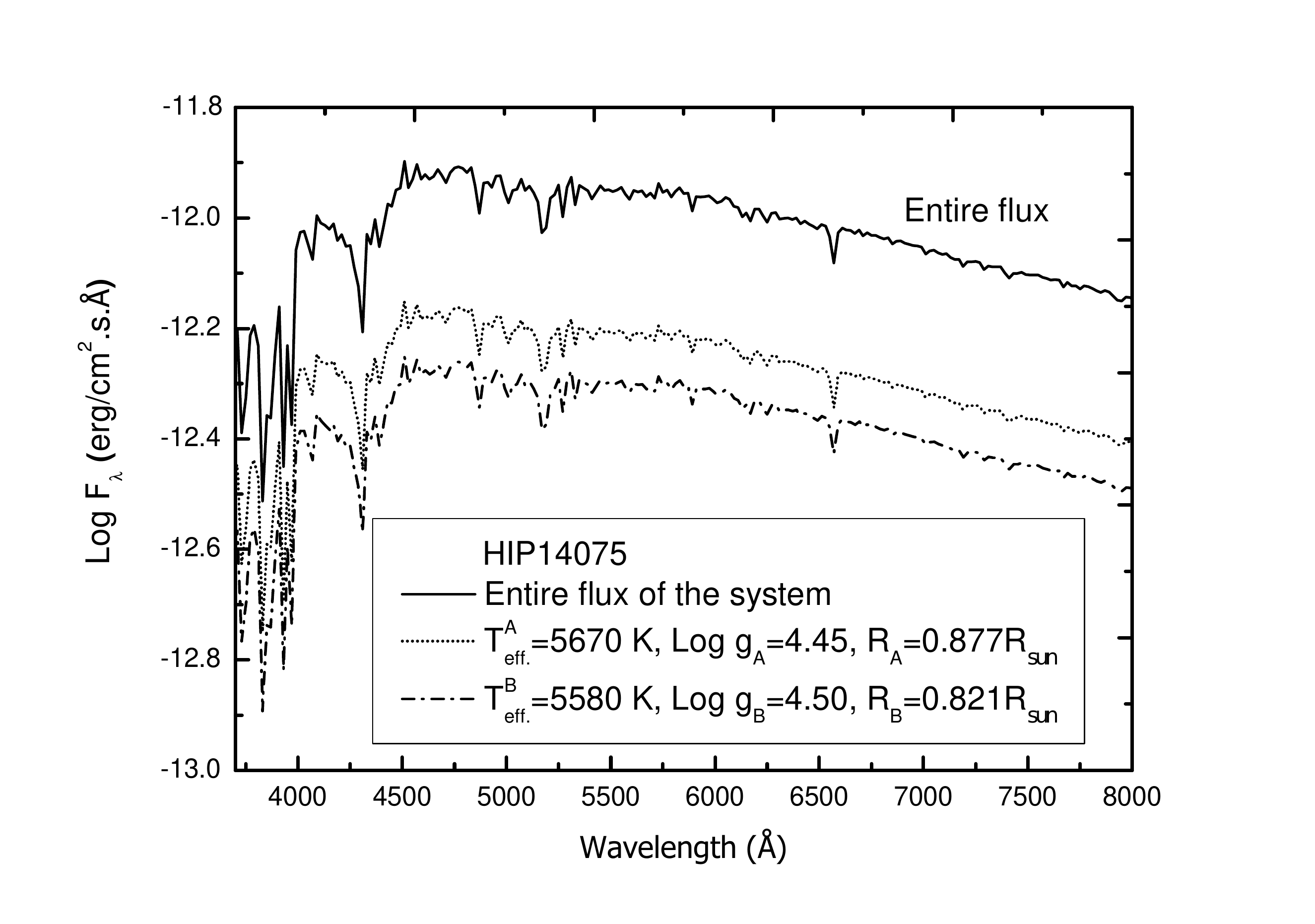}
    	\caption{The entire and individual synthetic SED of the system HIP 14075 using Kurucz blanketed models \citep{Kurucz} (ATLAS9) at distance of $ 60.17\pm0.001$ pc.\label{fig1}}
    \end{figure*}

\begin{figure*}
    	\centering
    	\includegraphics[angle=0,width=12cm]{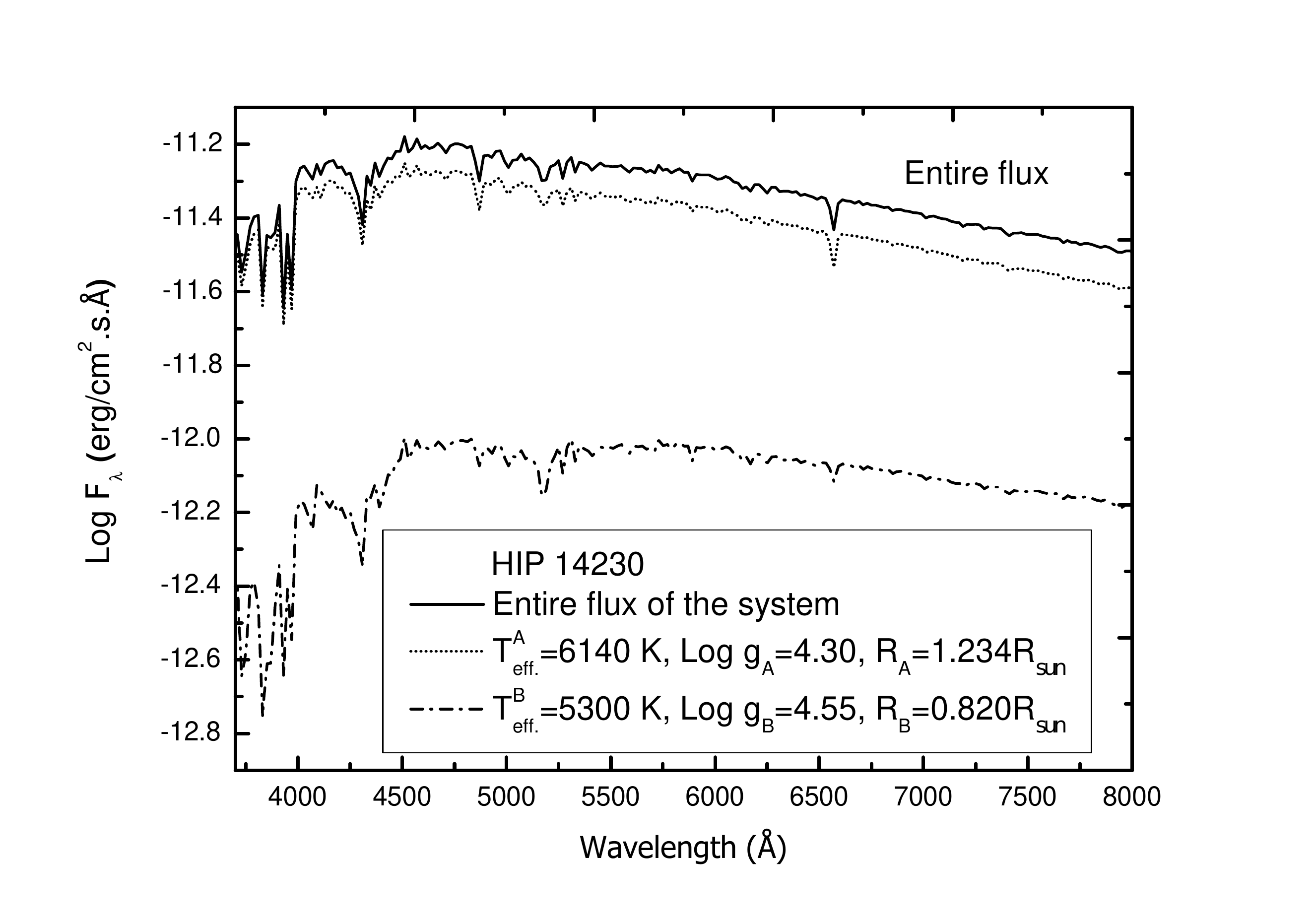}
    	\caption{Same as Fig.\ref{fig1} but here for the HIP\,14230 system at distance of $37.99\pm0.001$ pc .\label{fig2}}
    \end{figure*}

	Figs.~\ref{fig1}\& \ref{fig2} show the entire and individual synthetic spectral energy distributions of the old close binary systems, HIP 14075 and HIP 14230, respectively based on the calculated stellar parameters and on the Gaia parallaxes \citep{2018yCat.1345....0G}.		
    
	 The results of the apparent magnitudes $m_v$ from synthetic photometry are found to be completely similar to those from the observed photometry for both binaries. At the same time, the difference between synthetic and observed valuse of magnitudes and colours indices in the different photometric systems for both binaries is less than 2.1 $\sigma$. The agreement between these values indicates an accuracy of the method and an indication for the reliability of the calculated stellar parameters of the systems.
		
\begin{figure*}
  	\centering
  	\includegraphics[width=1.5\columnwidth]{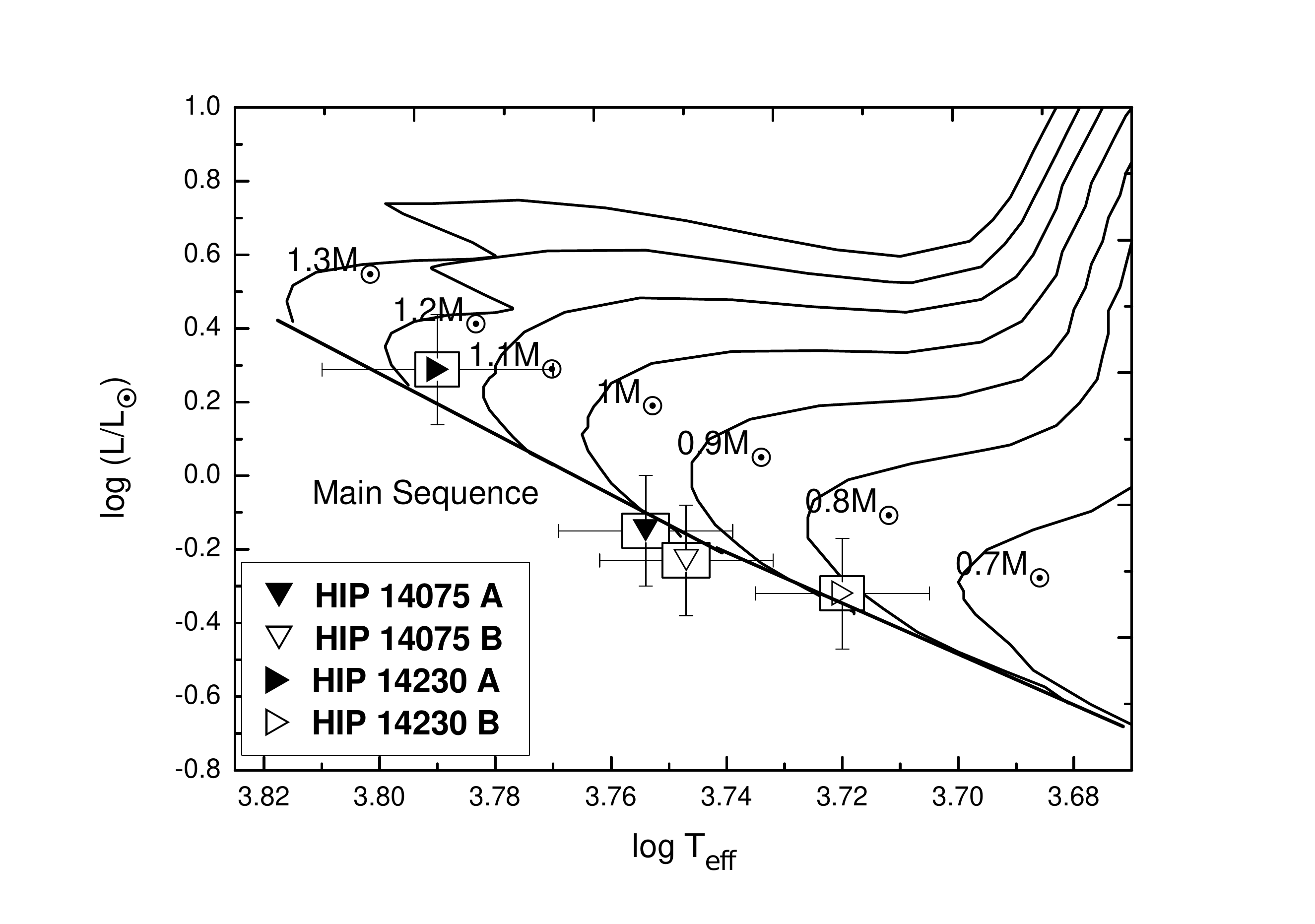}
  	\caption{The evolutionary tracks of both components of HIP\,14075 and of HIP\,14230 on the H-R diagram of masses ( 0.7, 0.8,...., 1.3 $\rm\,\mathcal{M_\odot}$). The evolutionary tracks were taken from~ \cite{Girardi2000b}.} \label{a25}
  \end{figure*}

  Fig.~\ref{a25} shows the components of both HIP 14075 and HIP 14230 binary systems on the evolutionary tracks of \cite{Girardi2000b} which belong to the main-sequence stars.
  
   \begin{figure*}
   	\centering
   	\includegraphics[width=1.5\columnwidth]{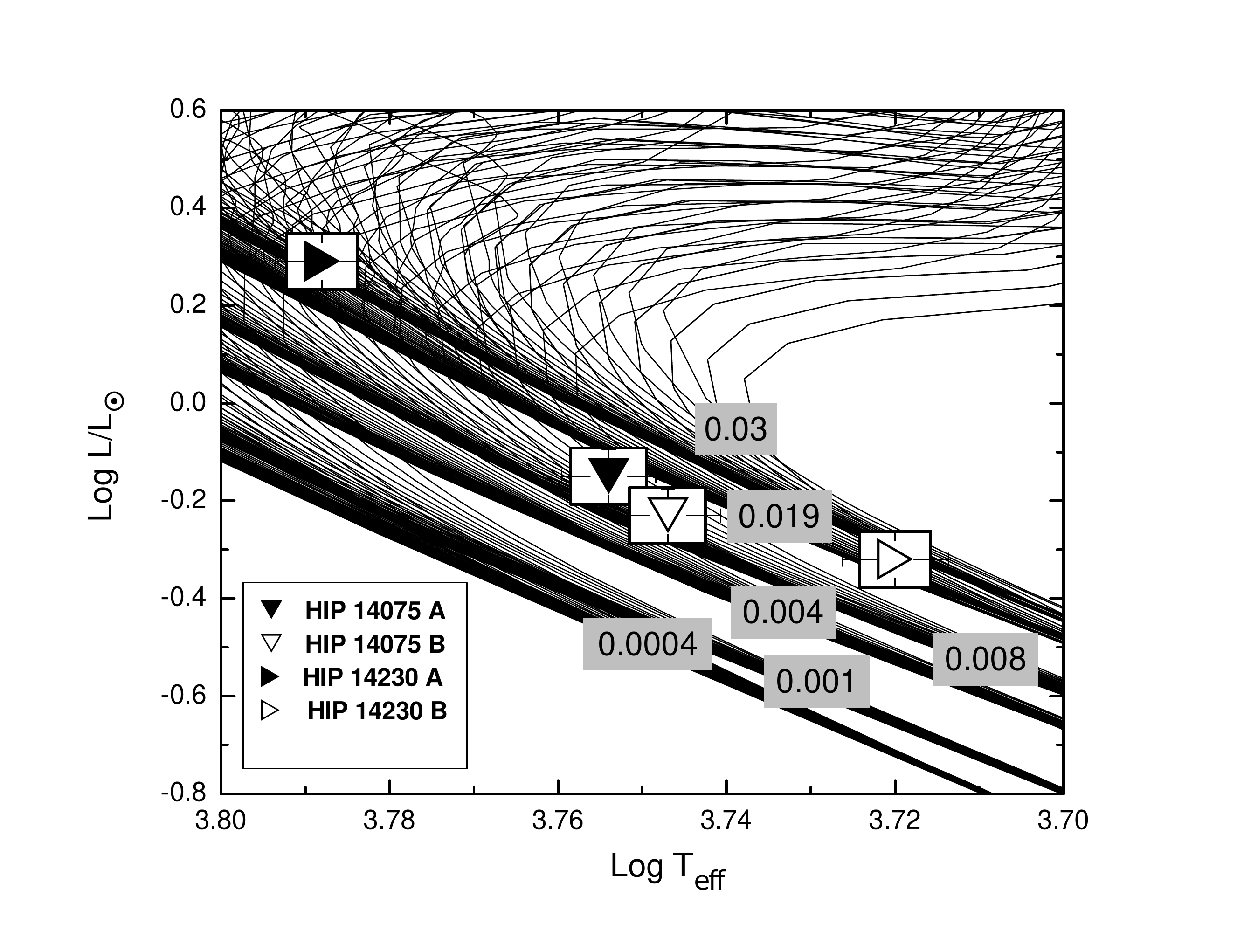}
   	\caption{The isochrones for both components of HIP\,14075 and of HIP\,14230 on the H-R diagram for low- and intermediate-mass: from $0.15 $ to $7.0 \rm\,\mathcal{M_\odot}$ stars of
   		different metallicities (from Z=0.0004 to 0.03). The isochrones were taken from~\cite{Girardi2000a}.} \label{a26}
   \end{figure*} 
  
  The lack of knowledge of the orbital inclination $i$ and  another orbital solutions of the systems prevents a direct determination of the component masses from the mass function $f(\mathcal{M})$ derived from the spectroscopic orbits. Thus, so as to estimate the stellar masses and ages of the systems, we used \cite{Girardi2000b}'s theoretical H-R diagram with evolution tracks, and the isochrones given by \cite{Girardi2000a}, respectively.
  
  Fig.~\ref{a25} shows the positions of the masses of HIP 14075 and HIP 14230 on the H-R evolutionary tracks. The corresponding
  mass values of the system HIP 14075 are $\mathcal{M}^A=0.99\pm0.19\, \mathcal{M_\odot}$, $\mathcal{M}^B=0.96\pm0.15\, \mathcal{M_\odot}$, with spectral types G4.5 for the primary component and G6.5 for the secondary component,  while the corresponding
  mass values of the system HIP 14230 are $\mathcal{M}^A=1.18\pm0.22\, \mathcal{M_\odot}$, $\mathcal{M}^B=0.84\pm0.12\, \mathcal{M_\odot}$, with spectral types F8 for the primary component and K0 for the secondary component.
  
  In comparison our results with previous results in terms of system masses, in case of HIP\,14075, the system mass is much accurate than that of \cite{Balega2005}, whilst in case of HIP\,14230 there is a significant change to that from  \cite{Balega2005}. Gaia parallaxes of the systems have improved the masses of each system. 

\begin{figure*}
  	\centering
  	\includegraphics[width=1.5\columnwidth]{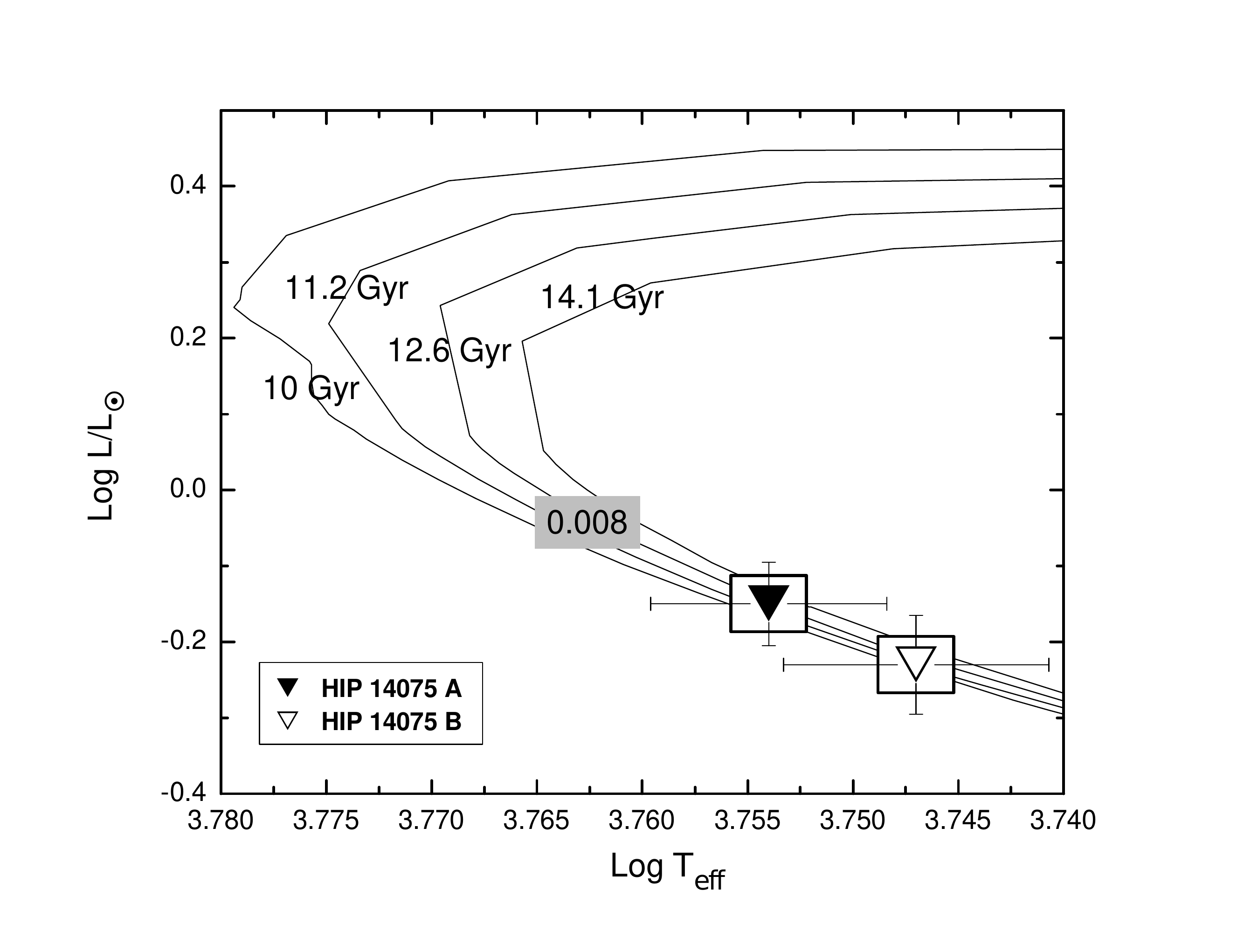}
  	\caption{The isochrones for both components of HIP\,14075 on the H-R diagram for low- and intermediate-mass: from $0.15 $ to $0.99 \rm\,\mathcal{M_\odot}$ , and for the compositions [Z=0.008, Y=0.25] stars. The isochrones were taken from~\cite{Girardi2000a}.} \label{a27}
  \end{figure*}

\begin{figure*}
  	\centering
  	\includegraphics[width=1.5\columnwidth]{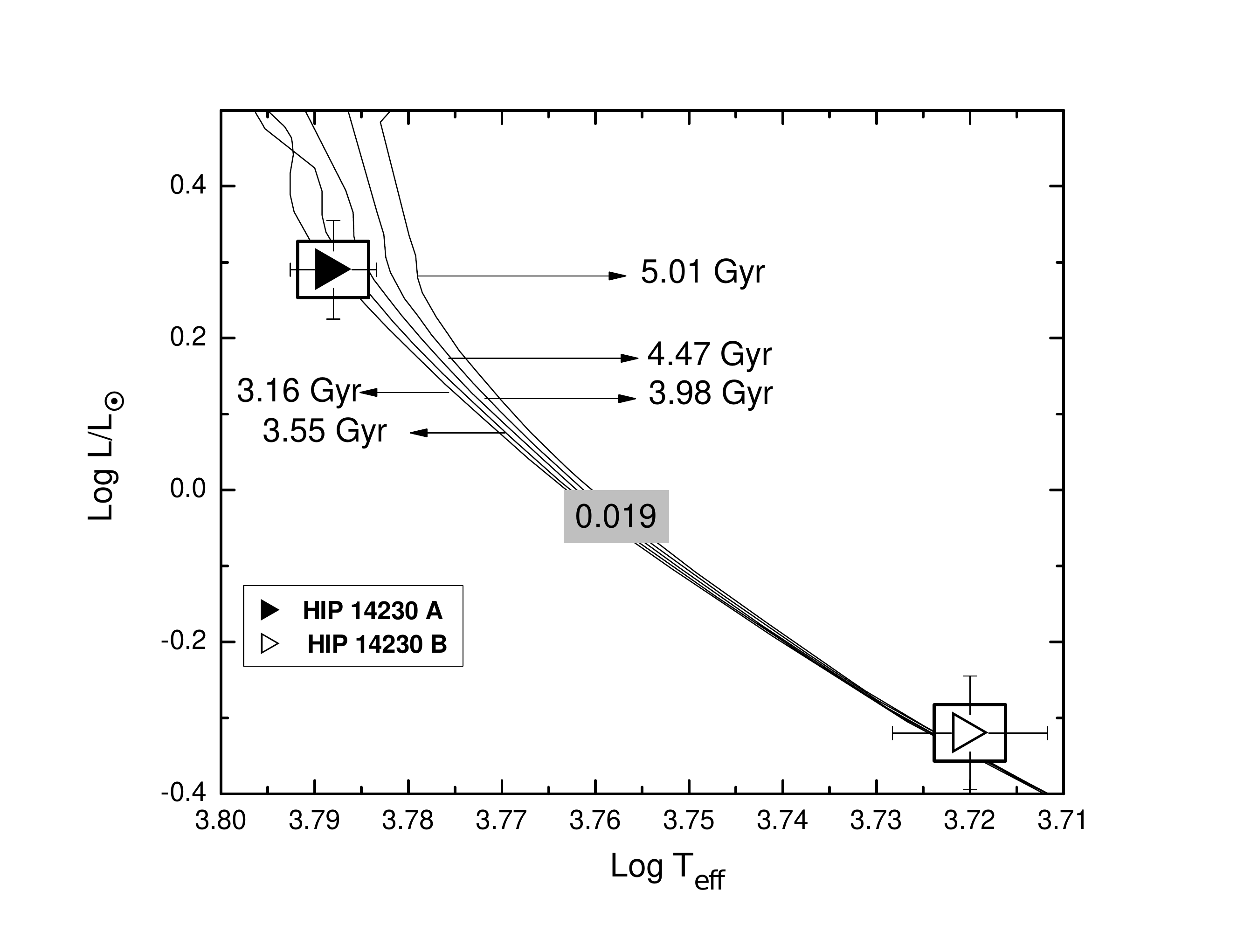}
  	\caption{The isochrones for both components of HIP\,14230 on the H-R diagram for low- and intermediate-mass: from $0.15 $ to $1.33 \rm\,\mathcal{M_\odot}$ , and for the compositions [Z=0.019, Y=0.273] stars. The isochrones were taken from~\cite{Girardi2000a}.} \label{a28}
  \end{figure*}

  Since the analysis of the HIP\,14075 and HIP\,14230 systems revealed that both systems are on the main sequence and the systems appear to be roughly coeval. As a result, our analysis of HIP\,14075 shows that the age of the system is around $11.5\pm2.0$\,Gyr, and the solar composition is [Z = 0.008, Y = 0.25] \citep{Girardi2000a} (Figs~\ref{a26}\&~\ref{a27}), while the age of HIP\,14230 is around  $3.5\pm 1.5$\,Gyr and the solar composition is [Z = 0.019, Y = 0.273]~\citep{Girardi2000a} (Figs~\ref{a26}\&~\ref{a28}). This analysis leads us to adopt
  the fragmentation process for the formation of such systems, 
  where \cite{1994MNRAS.269..837B} concludes that fragmentation
  of rotating disk around an incipient central protostar is possible,
  as long as there is continuing infall, and \cite{2001IAUS..200.....Z} pointed out that hierarchical fragmentation during
  rotational collapse has been invoked to produce binaries and
  multiple systems.
  

\section{Conclusions}\label{5}

We have presented the accurate stellar parameters of two old close binary systems, HIP\,14075 and  HIP 14230 by using synthetic photometric analysis. The synthetic SED of the two systems have been constructed using Al-Wardat's method and the results of the synthetic photometry have been depended on the results of Al-Wardat's  complex method. 

The present analyses show that the binaries HIP 14075 and HIP 14230 belong to a class of the old close binary systems ($\approx >$ 4 Gyr). The results from synthetic photometry are found to be similar to those from the observed ones, which revealed the accuracy of the used method and led to estimate the best stellar parameters for both systems.

The positions of the individual components of the systems using the evolutionary tracks and isochrones proved that the components of HIP 14075 and HIP 14230 belong to main sequence systems. The spectral types were estimated as G4.5 and G6.5 for the primary and secondary components, respectively for HIP\,14075, and as F8 and K0 for the primary and secondary components, respectively for HIP\,14230.

	
	\section*{Acknowledgements}
We thank the anonymous referee whose comments improved the paper.
This research has much more  made use of SAO/NASA, SIMBAD database, Fourth Catalog of Interferometric Measurements of Binary Stars, IPAC data systems and  CHORIZOS  code of photometric
	and spectrophotometric data analysis .

\nocite{*}

\end{document}